\documentclass[11pt,reqno]{amsart}
\usepackage[utf8]{inputenc}
\usepackage{amssymb,mathrsfs,color,bbold,comment,enumerate,xcolor}
\usepackage[colorlinks=true,urlcolor=blue,citecolor=magenta,linkcolor=blue]{hyperref}
\numberwithin{equation}{section}
\usepackage[all]{xy}
\usepackage{color}
\usepackage{amsaddr}
\usepackage{chngcntr}
\usepackage{apptools}
\usepackage[onehalfspacing]{setspace}
\AtAppendix{\counterwithin{theorem}{section}}
\AtAppendix{\counterwithin{equation}{section}}

\makeatletter
\renewcommand{\email}[2][]{
\ifx\emails\@empty\relax\else{\g@addto@macro\emails{,\space}}\fi
\@ifnotempty{#1}{\g@addto@macro\emails{\textrm{(#1)}\space}}
\g@addto@macro\emails{#2}
}
\makeatother

\makeatletter
\def\namedlabel#1#2{\begingroup
    #2%
    \def\@currentlabel{#2}%
    \phantomsection\label{#1}\endgroup
}
\makeatother

\theoremstyle{plain}
\newtheorem{theorem}{Theorem}[section]

\newtheorem*{theorem*}{Theorem}

\newtheorem{proposition}[theorem]{Proposition}

\newtheorem{lemma}[theorem]{Lemma}

\theoremstyle{definition}
\newtheorem{assumption}[theorem]{Assumption}
\newtheorem{remark}[theorem]{Remark}
\newtheorem{example}[theorem]{Example}

\DeclareMathOperator{\ES}{\rm ES}

\newcommand{\probp}{{\mathbb P}}
\newcommand{\probq}{{\mathbb Q}}

\newcommand{\R}{\mathbb{R}}

\newcommand{\E}{\mathbb{E}}
\newcommand{\tn}{\textnormal}

\newcommand{\ind}{\mathbf{1}}

\renewcommand{\P}{\mathbb{P}}

\newcommand{\N}{\mathbb{N}}
\newcommand{\dom}{\textnormal{dom}}

\newcommand{\om}{\omega}

\newcommand{\CF}{\mathcal F}

\newcommand{\CA}{\mathcal A}
\newcommand{\CG}{\mathcal G}

\newcommand{\CX}{\mathcal X}

\newcommand{\Q}{\mathbb{Q}}
\newcommand{\CC}{{\mathcal C}}
\newcommand{\cC}{\mathcal C}

\newcommand{\CL}{\mathcal L}

\newcommand{\CQ}{\mathcal Q}
\newcommand{\CH}{\mathcal H}

\newcommand{\ph}{\varphi}

\newcommand{\cX}{{\mathcal X}}

\newcommand{\Max}{\mathbf{Max}}

\title[Law-invariant functionals that collapse to the mean:
Beyond convexity]{Law-invariant functionals that collapse to the mean:\\
Beyond convexity}

\author[F.-B.~Liebrich]{Felix-Benedikt Liebrich}
\address{Institute of Actuarial and Financial Mathematics \& House of Insurance,\\
Leibniz University Hannover, Germany}
\email{felix.liebrich@insurance.uni-hannover.de}

\author[C.~Munari]{Cosimo Munari}
\address{Center for Finance and Insurance \& Swiss Finance Institute,\\
University of Zurich, Switzerland}
\email{cosimo.munari@bf.uzh.ch}

\date{July 13, 2021}

\begin{document}

\parindent 0em \noindent

\begin{abstract}
We establish general ``collapse to the mean'' principles that provide conditions under which a law-invariant functional reduces to an expectation. In the convex setting, we retrieve and sharpen known results from the literature. However, our results also apply beyond the convex setting. We illustrate this by providing a complete account of the ``collapse to the mean'' for quasiconvex functionals. In the special cases of consistent risk measures and Choquet integrals, we can even dispense with quasiconvexity. In addition, we relate the ``collapse to the mean" to the study of solutions of a broad class of optimisation problems with law-invariant objectives that appear in mathematical finance, insurance, and economics. We show that the corresponding quantile formulations studied in the literature are sometimes illegitimate and require further analysis.

\medskip

\end{abstract}

\maketitle



\section{Introduction}

The expression ``collapse to the mean'' refers to a variety of results about law-invariant functionals defined on spaces of random variables. The common thread of such results lies in the fundamental tension existing between law invariance and suitable ``linearity'' properties (linearity, affinity, translation invariance). In the context of mathematical finance, insurance, or economics, a random variable typically models the future unknown value of a given financial or economic variable of interest (the payoff of an asset, the return on a portfolio of assets, the net worth of an agent, the capital level of a financial institution).
The functional under consideration models the ``value'' of said variable (a price, a risk measure, a utility index, a capital requirement). In this context, the assumption of law invariance posits that ``value'' is only sensitive to the distribution of the underlying variables with respect to a given reference probability measure, so that statistical tools may be used to perform estimation in concrete situations.
The assumption of ``linearity'' typically captures the presence of a frictionless determinant of ``value'' (a riskless investment opportunity, a liquidly traded asset without transaction costs). As the term suggests, the ``collapse to the mean'' is concerned with properties under which the only functionals that are simultaneously law invariant and ``linear'' are expectations or, more generally, functions of the expectation (with respect to the reference probability measure). These results are an important litmus test because functionals that are fully determined by expectation typically fail to capture ``value'' in an adequate risk-sensitive way. Avoiding an inadequate representation of ``value'' would thus force a choice between law invariance and other properties that are often desirable on their own merits.

To our knowledge, the earliest ``collapse to the mean'' is recorded in \cite{Castagnoli}, which proves that the expectation is the only law-invariant Choquet integral defined on the space of bounded random variables that is convex and linear along a nonconstant random variable. This result has natural applications to the literature on Choquet pricing. It shows that the combination of law invariance --- a common postulate in insurance pricing --- and the existence of a frictionless risky traded asset is only compatible with frictionless markets where prices are determined by expectation (with respect to the physical probability measure) and where, as a consequence, obvious arbitrage opportunities arise. The collapse for Choquet integrals was later extended, again in a bounded setting, to general cash-additive functionals in \cite{Frittelli}. Further extensions beyond the bounded setting but retaining the convexity assumption have recently been obtained in \cite{Collapse}, to which we refer for further information. In the recent working paper~\cite{WangWu}, the authors show that a functional defined on bounded random variables is a function of the expectation if and only if it is dependence neutral, i.e., the functional applied to a sum of random variables only depends on their marginal distributions. Notably, \cite{WangWu} does not impose convexity assumptions. A collapse to the mean for conditionally convex maps has been recently obtained in \cite{Delbaen2021}.

The goal of this paper is to present general formulations of the ``collapse to the mean'' that both extend the known results from the literature and can be applied beyond the world of convex functionals. The general ``collapse to the mean'' principle is stated in Theorem~\ref{theo:meta}, which in turn is derived from a sharp version of the Fr\'echet-Hoeffding bounds recorded in Lemma~\ref{lem:interior}. A complementary geometric version of the general principle is stated in Proposition~\ref{prop: law invariance recession cone collapse}. We illustrate the versatility of these tools in five case studies.

\smallskip

{\bf Collapse for convex functionals}. In Section~\ref{sect: convex}, we revisit the known ``collapse to the mean'' for convex functionals. We provide two versions under the assumption that the underlying functional is translation invariant along a nonconstant random variable, see Theorem~\ref{th: zero collapse convex} and Theorem~\ref{thm:affine}. If the random variable has zero expectation, the functional collapses to a function of the expectation. Otherwise, it collapses to a specific function, namely an affine function, of the expectation. This confirms the results in \cite{Collapse,Castagnoli,Frittelli}. In addition, we provide new dual characterizations of the collapse in terms of weaker translation invariance properties and conjugate functions.

\smallskip

{\bf Collapse for quasiconvex functionals}. In Section~\ref{sect: quasiconvex} we take up the study of quasiconvex functionals. This is an important extension in view of the economic interpretation of quasiconvexity, which is a more elementary mathematical formulation of the diversification principle; see, e.g., \cite{Diversification,Drapeau,Gao1,Gao2,Liebrich,Rosazza} in a risk measure context. We extend both convex versions of the collapse, see Theorem~\ref{thm:main} and Theorem~\ref{thm:extquasiconv}, by means of the aforementioned sharp Fr\'echet-Hoeffding bounds. Moreover, we demonstrate sharpness of our results.

\smallskip

{\bf Collapse for consistent risk measures}. In Section~\ref{sect: consistent risk measures}, we focus on cash-additive functionals that are monotonic with respect to second-order stochastic dominance. This class of risk measures is named ``consistent" in \cite{Consistent} and contains the family of law-invariant convex risk measures, but also functionals that are neither convex nor quasiconvex. The literature on the connection between risk measures and stochastic dominance is rich; see, e.g., \cite{BauerleMuller,DeGiorgi,Leitner,Semideviation,Dual Semideviation}. The collapse for consistent risk measures is recorded in Theorem~\ref{thm:consistent}, which is based again on the sharp version of the Fr\'echet-Hoeffding bounds.

\smallskip

{\bf Collapse for Choquet integrals}. In Section~\ref{sect: Choquet}, we take one further step beyond convexity and consider Choquet integrals associated with a variety of different law-invariant capacities. In the case of submodular capacities, the Choquet integral is convex and a related collapse to the mean was obtained in \cite{Castagnoli}. We go beyond submodular capacities and consider the case of coherent as well as Jaffray-Philippe capacities. The corresponding Choquet integrals are neither convex nor quasiconvex and play a natural role in decision theory under ambiguity; see, e.g., \cite{Chateauneuf,JP,Ravanelli}. For a review of capacities and Choquet integrals, we refer to \cite{MariMont} and the references therein. In Theorem~\ref{thm:Choquet} we use the sharp Fr\'echet-Hoeffding bounds to derive a collapse result for this general class of Choquet integrals.

\smallskip

{\bf Collapse in optimisation problems}. In Section~\ref{sect: Optimisation} we focus on a general optimisation problem that encompasses a variety of important problems in economics, finance, and insurance, including the maximisation of expected investment returns or expected utility from terminal wealth (von Neumann-Morgenstern utility, rank-dependent utility, Yaari utility, S-shaped utility from prospect theory). More precisely, we study the maximisation of a general law-invariant objective subject to a general law-invariant constraint and a ``budget'' constraint expressed in terms of a ``pricing density''. A common intuition for such optimisation problems is that, \textit{if} a solution exists, then all or some of these solutions have to be \textit{antimonotone} with the pricing density. This allows to reduce the original problem to an optimisation problem involving quantile functions, which is substantially simpler and for which solution techniques are
 available; see, e.g., \cite{Burgert,Carlier,HeZhou,RueschendorfVanduffel,Schied,Xu,Xu2}. We provide a slight improvement over the existing results---see in particular \cite{Xu}---by establishing more general sufficient conditions for the existence of antimonotone solutions. In particular, we highlight some conditions that are often omitted in the literature. In addition, we conduct a careful analysis showing that our result is sharp in the sense that, if any of the conditions is removed, the validity of the result forces the budget constraint to ``collapse to the mean'': The pricing density is necessarily constant, and the corresponding pricing rule reduces to the expectation with respect to the physical probability measure. This points to an issue in the literature, where the reduction to a quantile formulation is sometimes invoked even though some of the aforementioned conditions are not satisfied. In this situation, the reduction might be illegitimate unless extra analysis of the
  {\em specific} structure of the problem is carried over.

\smallskip

The paper is organised as follows. In Section~\ref{sect: setting} we describe the underlying setting and introduce the necessary notation. In Section~\ref{sect: key tools} we record our main tool, namely the sharp Fr\'echet-Hoeffding bounds. In Section~\ref{sect: general principle} we state the general ``collapse to the mean'' principle and establish a useful geometric counterpart for convex sets. In Section~\ref{sect: applications} we provide a range of applications to convex and quasiconvex functionals, consistent risk measures, and Choquet integrals. In addition, we discuss a general optimisation problem involving law invariance, provide a result about optimal solutions, and show what can go wrong when passing to its quantile formulation. Appendix~\ref{appendix} provides a proof of Lemma~\ref{lem:interior}.


\section{Setting and notation}
\label{sect: setting}

Let $(\Omega,\CF,\P)$ be an atomless probability space. A Borel measurable function $X\colon\Omega\to\R$ is called a random variable. By $L^0$ we denote the set of equivalence classes of random variables with respect to almost-sure equality under $\probp$. As is customary, we do not explicitly distinguish between an element of $L^0$ and any of its representatives. In particular, the elements of $\R$ are naturally identified with random variables that are almost-surely constant under $\probp$. For two random variables $X,Y\in L^0$ we write $X\sim Y$ whenever $X$ and $Y$ have the same law with respect to $\probp$, i.e., the probability measures $\P\circ X^{-1}$ and $\P\circ Y^{-1}$ on the real line agree. The expectation operator under $\probp$ is denoted by $\E[\cdot]$. The standard Lebesgue spaces are denoted by $L^p$ for $p\in[1,\infty]$. We say that a set $\cX\subset L^0$ is {\em law invariant} if $X\in\cX$ for every $X\in L^0$ such that $X\sim Y$ for some $Y\in\cX$.

\begin{assumption}\label{XX^*}
We denote by $(\cX,\cX^\ast)$ a pair of law-invariant vector subspaces of $L^1$ containing $L^\infty$. We assume that $XY\in L^1$ for all $X\in\cX$ and $Y\in\cX^\ast$ and denote by $\sigma(\cX,\cX^\ast)$ the weakest linear topology on $\cX$ with respect to which, for every $Y\in\cX^\ast$, the linear functional on $\cX$ given by $X\mapsto\E[XY]$ is continuous.\footnote{~Note that, as $\cX$ and $\cX^\ast$ contain $L^\infty$ by assumption, the pairing on $\cX\times\cX^\ast$ given by $(X,Y)\mapsto\E[XY]$ is separating. In particular, when equipped with the topology $\sigma(\cX,\cX^\ast)$, the space $\cX$ is a locally convex Hausdorff topological vector space.}
\end{assumption}

We say that a (nonempty) set $\cC\subset\cX$ is {\em convex} if it contains the convex combination of any of its elements, and {\em $\sigma(\cX,\cX^\ast)$-closed} if it contains the limit of any $\sigma(\cX,\cX^\ast)$-convergent net of its elements. The {\em (upper) support functional} of $\cC$ is the map $\sigma_\cC\colon\cX^\ast\to[-\infty,\infty]$ given by
\[
\sigma_\cC(Y) := \sup_{X\in\cC}\E[XY].
\]
Throughout the paper we focus on functionals $\ph\colon\cX\to[-\infty,\infty]$. The {\em domain} of $\ph$ is
\[
\dom(\ph) := \{X\in\cX \,; \ \ph(X)\in\R\}.
\]
We say that $\ph$ is {\em proper} if $\dom(\ph)$ is nonempty. Moreover, the functional $\ph$ is called:
\begin{enumerate}[(1)]
  \item {\em convex} if for all $X,Y\in\cX$ and $\lambda\in[0,1]$,
\[
\ph(\lambda X+(1-\lambda)Y)\leq\lambda\ph(X)+(1-\lambda)\ph(Y).
\]
  \item {\em quasiconvex} if for all $X,Y\in\cX$ and $\lambda\in[0,1]$,
\[
\ph(\lambda X+(1-\lambda)Y)\leq\max\{\ph(X),\ph(Y)\}.\footnote{~Equivalently, for every $m\in\R$ the lower level set $\{X\in\CX\,;\ \ph(X)\le m\}$ is convex.}
\]
  \item {\em $\sigma(\cX,\cX^\ast)$-lower semicontinuous} if for all nets $(X_\alpha)\subset\cX$ and $X\in\cX$,
\[
X_\alpha\xrightarrow{\sigma(\cX,\cX^\ast)}X \ \implies \ \ph(X)\leq\liminf_{\alpha}\ph(X_\alpha).\footnote{~Equivalently, for every $m\in\R$ the lower level set $\{X\in\CX\,;\ \ph(X)\le m\}$ is $\sigma(\cX,\cX^\ast)$-closed.}
\]
  \item {\em law invariant} if for all $X,Y\in\cX$,
\[
X\sim Y \ \implies \ \ph(X)=\ph(Y)
\]
\item {\em expectation invariant} if for all $X,Y\in\cX$,
\[
\E[X]=\E[Y] \ \implies \ \ph(X)=\ph(Y).\footnote{~Equivalently, for every $X\in\cX$ we have $\ph(X)=\ph(\E[X])$.}
\]
\item an {\em affine function of the expectation} if there exist $a,b\in\R$ such that, for every $X\in\cX$,
\[
\ph(X)=a\E[X]+b.
\]
\end{enumerate}
The {\em conjugate} of $\ph$ is the functional $\ph^\ast\colon\cX^\ast\to[-\infty,\infty]$ given by
\[
\ph^\ast(Y) := \sup_{X\in\cX}\{\E[XY]-\ph(X)\}.
\]
The next lemma records the well-known dual representation of convex closed sets and convex lower-semicontinuous functionals, which are direct consequences of the Hahn-Banach theorem; see, e.g., \cite[Theorem 1.1.9, Theorem 2.3.3]{Zalinescu}.

\begin{proposition}
\label{prop: dual representation}
Let $\cC\subset\cX$ be convex and $\sigma(\cX,\cX^\ast)$-closed. Then,
\[
\cC = \bigcap_{Y\in\cX^\ast}\{X\in\cX \,; \ \E[XY]\leq\sigma_\cC(Y)\}.
\]
Let $\ph\colon\cX\to(-\infty,\infty]$ be proper, convex, and $\sigma(\cX,\cX^\ast)$-lower semicontinuous. Then,
\[
\ph(X) = \sup_{Y\in\cX^\ast}\{\E[XY]-\ph^\ast(Y)\}, \ \ \ X\in\cX.
\]
\end{proposition}

One of the guiding threads of this paper is the fact that for many classes of functionals there is a fundamental tension between law invariance and suitable ``linearity'' properties. A prominent class consists of quasiconvex functionals. In this case, the property of law invariance is equivalent to other well-known properties such as dilatation monotonicity and Schur convexity, to which our results therefore naturally apply. We refer to \cite[Theorem 3.6, Proposition 5.6]{Bellini} for a proof in our general setting.

\begin{proposition}
\label{prop: equivalence law invariance}
Let $\ph\colon\cX\to(-\infty,\infty]$ be proper, quasiconvex, and $\sigma(\cX,\cX^\ast)$-lower semicontinuous. Then, the following statements are equivalent:
\begin{enumerate}[(i)]
    \item $\ph$ is law invariant.
    \item $\ph$ is dilatation monotone, i.e., for every $X\in\cX$ and every $\sigma$-field $\CG\subset\CF$,
\[
\E[X\vert\CG]\in\cX \ \implies \ \ph(X)\geq\ph(\E[X\vert\CG]).
\]
    \item $\ph$ is Schur convex, i.e., for all $X,Y\in\cX$,
\[
\mbox{$\E[f(X)]\geq\E[f(Y)]$ for every convex $f\colon\R\to\R$} \ \implies \ \ph(X)\geq\ph(Y).
\]
\end{enumerate}
\end{proposition}


\section{The key tool: Sharp Fr\'{e}chet-Hoeffding bounds}
\label{sect: key tools}

In this brief section we record the main tool that is needed to establish our ``collapse to the mean'' results, which consists of a sharp formulation of the well-known Fr\'{e}chet-Hoeffding bounds. For any random variable $X\in L^0$ we denote by $q_X$ a {\em fixed} quantile function of $X$, i.e., a function $q_X\colon(0,1)\to\R$ satisfying for every $s\in(0,1)$
\[
\inf\{x\in\R \,; \ \probp(X\leq x)\geq s\} \leq q_X(s) \leq \inf\{x\in\R \,; \ \probp(X\leq x)>s\}.
\]
As the distribution function of $X$ has at most countably many discontinuity points, any two quantile functions of $X$ coincide almost surely with respect to the Lebesgue measure on $(0,1)$. For $X,Y\in L^0$ we say that $X$ and $Y$ are {\em comonotone} if for all $x,y\in\R$,
\[
\probp(X\leq x,Y\leq y) = \min\{\probp(X\leq x),\probp(Y\leq y)\}.\footnote{~Equivalently, there are nondecreasing functions $f,g\colon\R\to\R$ and $Z\in L^0$ with $X=f(Z)$ and $Y=g(Z)$; cf.\ \cite[Proposition 4.5]{Denneberg}.}
\]
Similarly, we say that $X$ and $Y$ are {\em antimonotone} if for all $x,y\in\R$,
\[
\probp(X\leq x,Y\leq y) = \max\{\probp(X\leq x)+\probp(Y\leq y)-1,0\}.\footnote{~Equivalently, $X$ and $-Y$ or $-X$ and $Y$ are comonotone.}
\]

In the proof of the sharp version of the Fr\'{e}chet-Hoeffding bounds and in the sequel, we will repeatedly use the fact that, by nonatomicity, for all $X,Y\in L^0$ we can always find $X'\sim X$ and $Y'\sim Y$ such that $X'$ and $Y'$ are comonotone. The analogue for anticomonotonicity holds as well. In fact, we have the following stronger result.

\begin{lemma}
\label{lem:existence}
For all $X\in\cX$ and $Y\in\cX^\ast$ there exist $X',X''\sim X$ such that $X'$ and $Y$ are comonotone and $X''$ and $Y$ are antimonotone.
\end{lemma}
\begin{proof}
By nonatomicity, we find a uniform random variable $U\in L^0$ on $(0,1)$ such that $Y=q_Y(U)$; see, e.g., \cite[Theorem 5]{Xu}. It is then enough to take $X'=q_X(U)$ and $X''=q_X(1-U)$.
\end{proof}

The next result connecting the range of special integrals and quantile functions builds on early work by Fr\'{e}chet and Hoeffding on joint distribution functions (see \cite{Burgert}) and Chebyshev, Hardy, and Littlewood on rearrangement inequalities (see \cite{Luxemburg}). Its general formulation in our setting is essentially due to Luxemburg; see \cite[Theorem 9.1]{Luxemburg}. However, as the statements found in the literature contain only portions of the statement we need, we provide a complete proof in our general framework in Appendix~\ref{appendix}.

\begin{lemma}
\label{lem:interior}
For all $X\in\cX$ and $Y\in\cX^\ast$ the functions
\begin{center}$(0,1)\ni s\mapsto q_X(s)q_Y(s)$ \ and \ $(0,1)\ni s\mapsto q_X(s)q_Y(1-s)$\end{center}
are both Lebesgue integrable on $(0,1)$ and
\begin{equation}
\label{eq:interval}
\min_{X'\sim X}\E[X'Y]=\int_0^1q_X(1-s)q_Y(s)ds, \ \ \ \max_{X'\sim X}\E[X'Y]=\int_0^1q_X(s)q_Y(s)ds.
\end{equation}
The minimum, respectively maximum, is attained by $X'\sim X$ if and only if $X'$ and $Y$ are antimonotone, respectively comonotone. Moreover, if both $X$ and $Y$ are nonconstant,
\begin{equation}
\label{eq: strict hardy littlewood}
\int_0^1q_X(1-s)q_Y(s)ds<\E[X]\E[Y]<\int_0^1q_X(s)q_Y(s)ds.
\end{equation}
\end{lemma}


\section{The general ``collapse to the mean'' principle}
\label{sect: general principle}

This section contains our prototype version of the ``collapse to the mean'', which will later be exploited to obtain a variety of results for specific classes of functionals. This general result shows that the expectation is, up to an affine transformation, the only linear and $\sigma(\cX,\cX^\ast)$-continuous functional that is dominated above by a law-invariant functional which fulfills a suitable local translation invariance property. It should be noted that the result holds for a general law-invariant functional without any additional property.

\begin{theorem}
\label{theo:meta}
Let $\ph\colon\CX\to(-\infty,\infty]$ be law invariant and satisfy $\dom(\ph)\cap \R\neq\emptyset$. Assume that for some $x\in\dom(\ph)\cap \R$ there exist $a\in\R$ and a nonconstant $Z\in\CX$ such that
\[
\ph(x+tZ) = \ph(x)+at,\quad t\in\R.
\]
Then, $\dom(\ph^*)\subset\R$. In particular, if there exist $c\in\R$ and $Y\in\CX^\ast$ such that
\[
\ph(X)\geq\E[XY]+c,\quad X\in\CX,
\]
then $Y$ must be constant.
\end{theorem}
\begin{proof}
If $\dom(\ph^*)=\emptyset$, the assertion trivially holds. Hence, suppose we can select $Y\in\dom(\ph^*)$. By an affine transformation of $\ph$, we can assume without loss of generality that $\ph^*(Y)=0$. For all $k\in \N$ and $Z'\sim Z$, we observe that
\[
\ph(x+kZ)-\ph(x) = \ph(x+kZ')-\ph(x) \geq k\E[Z'Y]+x\E[Y]-\ph(x).
\]
In the same vein,
\begin{align*}
\ph(x+kZ)-\ph(x) &= \ph(x)-\ph(x-kZ) \\
&= \ph(x)-\ph(x-kZ') \leq \ph(x)+k\E[Z'Y]-x\E[Y].
\end{align*}
As a result, for every $k\in\N$,
\[
\sup_{Z'\sim Z}\E[Z'Y]\le\frac{2\left(\ph(x)-x\E[Y]\right)}{k}+\inf_{Z'\sim Z}\E[Z'Y].
\]
Letting $k\to\infty$, we infer that
\[\sup_{Z'\sim Z}\E[Z'Y]=\inf_{Z'\sim Z}\E[Z'Y].\]
As $Z$ is nonconstant, Lemma~\ref{lem:interior} implies that $Y$ has to be constant.
\end{proof}

We complement the previous theorem with a geometrical counterpart about convex sets. Recall that the {\em recession cone} of a convex set $\cC\subset\cX$ is defined by
\[
\cC^\infty := \left\{X\in\cX \,; \ \{X\}+\cC\subset\cC\right\}.
\]
The recession cone of $\cC$ is the set of all directions of recession of the set $\cC$. Before stating the announced result, it is useful to highlight the following dual representation of the recession cone of a law-invariant set.

\begin{lemma}
\label{lem: dual representation recession cone}
Let $\cC\subset\cX$ be convex and  $\sigma(\cX,\cX^\ast)$-closed. Then,
\begin{equation}
\label{eq:first}
\cC^\infty = \bigcap_{Y\in\dom(\sigma_\cC)}\{X\in\cX \,; \ \E[XY]\leq 0\}.
\end{equation}
If $\CC$ is law invariant, then
\begin{equation}
\label{eq:second}
\CC^\infty=\bigcap_{Y\in\dom(\sigma_\cC)}\left\{X\in\cX \,; \ \int_0^1q_X(s)q_Y(s)ds\leq 0\right\}.
\end{equation}
In particular, $\CC^\infty$ is law invariant itself.
\end{lemma}
\begin{proof}
To show \eqref{eq:first}, fix an arbitrary $U\in\CC$. It follows from Proposition \ref{prop: dual representation} that
\begin{align*}
\CC^\infty&=\{X\in\CX \,; \  \forall\,k\in\N, \ U+kX\in\CC\}\\
&=\bigcap_{Y\in\dom(\sigma_\CC)}\{X\in\CX \,; \ \forall\,k\in\N, \ \E[(U+kX)Y]\le \sigma_\CC(Y)\}\\
&=\bigcap_{Y\in\dom(\sigma_\CC)}\{X\in\CX \,; \ \forall\,k\in\N, \ \E[XY]\le \tfrac 1 k(\sigma_\CC(Y)-\E[UY])\}\\
&=\bigcap_{Y\in\dom(\sigma_\CC)}\{X\in\CX \,; \ \E[XY]\le 0\}.
\end{align*}
To show \eqref{eq:second}, note that law invariance of $\cC$ together with Lemma~\ref{lem:interior} imply for every $Y\in\cX^\ast$
\[
\sigma_\cC(Y) = \sup_{X\in\cC}\E[XY] = \sup_{X\in\cC}\sup_{X'\sim X}\E[X'Y] = \sup_{X\in\cC}\int_0^1q_X(s)q_Y(s)ds.
\]
This shows that $\sigma_\cC$ is a law-invariant functional and, thus, $\dom(\sigma_\cC)$ is a law-invariant set. As a result, we infer from \eqref{eq:first} together with Lemma~\ref{lem:interior} that
\begin{align*}
\cC^\infty
&=
\bigcap_{Y\in\dom(\sigma_\cC)}\{X\in\cX \,; \ \E[XY]\leq0\} \\
&=
\bigcap_{Y\in\dom(\sigma_\cC)}\bigcap_{Y'\sim Y}\{X\in\cX \,; \ \E[XY']\leq0\} \\
&=
\big\{X\in\cX \,; \ \forall\,Y\in\dom(\sigma_\cC), \ \sup_{Y'\sim Y}\E[XY']\leq0\big\} \\
&=
\big\{X\in\cX \,; \ \forall\,Y\in\dom(\sigma_\cC), \ \int_0^1q_X(s)q_Y(s)ds\leq0\big\}.
\end{align*}
This representation clearly shows that $\CC^\infty$ is law invariant.
\end{proof}

We are now ready to prove the announced geometrical version of the ``collapse to the mean'', which generalises an earlier result formulated in \cite[Proposition 5.10]{LiebrichSharing} and provides a simpler proof. It shows that a convex and $\sigma(\cX,\cX^\ast)$-closed set that is law invariant and admits a nonzero direction of recession with zero expectation must be determined by expectation: Whether or not a random variable belongs to the set depends exclusively on its mean. In particular, the set must contain infinitely many affine spaces.

\begin{proposition}
\label{prop: law invariance recession cone collapse}
Let $\cC\subset\cX$ be convex,  $\sigma(\cX,\cX^\ast)$-closed, and law invariant. If there exists a nonzero $Z\in\cC^\infty$ such that $\E[Z]=0$, then $\dom(\sigma_\cC)\subset\R$ and
\begin{equation}\label{eq:bounds}
\cC = \{X\in\cX \,; \ -\sigma_\cC(-1)\leq\E[X]\leq\sigma_\cC(1)\}.
\end{equation}
\end{proposition}
\begin{proof}
Since $Z\in\cC^\infty$ by assumption, Lemma \ref{lem: dual representation recession cone} implies that, for every $Y\in\dom(\sigma_\cC)$,
\[
\int_0^1q_Z(s)q_Y(s)ds \leq 0.
\]
Note that $Z$ is nonconstant by assumption. If there existed a nonconstant $Y\in\dom(\sigma_\cC)$, then Lemma~\ref{lem:interior} would entail the impossible chain of inequalities
\[
0 = \E[Z]\E[Y] < \int_0^1q_Z(s)q_Y(s)ds \leq 0.
\]
This yields $\dom(\sigma_\cC)\subset\R$. By positive homogeneity of $\sigma_\CC$, Proposition \ref{prop: dual representation} implies
\[
\cC =
\bigcap_{Y\in\dom(\sigma_\cC)}\{X\in\cX \,; \ \E[XY]\leq\sigma_\cC(Y)\} =
\{X\in\cX \,; \ -\sigma_\cC(-1)\leq\E[X]\leq\sigma_\cC(1)\}.
\]
This delivers the desired claims and concludes the proof.
\end{proof}


\section{Applications}
\label{sect: applications}

\subsection{Collapse to the mean: The convex case}
\label{sect: convex}

As stated in the introduction, a variety of ``collapse to the mean'' results have been established in the literature for convex functionals. Early versions of the collapse to the mean were obtained in \cite{Castagnoli} for convex Choquet integrals and in \cite{Frittelli} for convex monetary risk measures. The focus of both papers was on bounded random variables. A general version of the collapse to the mean for convex functionals beyond the bounded setting has recently been established in \cite{Collapse}. To best appreciate the differences with the quasiconvex case, we devote this section to revisiting the most general results from the literature and complementing them with additional conditions.

We start by revisiting \cite[Theorem 4.7]{Collapse}. This result states that, under convexity and $\sigma(\cX,\cX^\ast)$-lower semicontinuity, a functional that is law invariant and affine (in particular, linear) along a nonconstant random variable with zero expectation must be, in our terminology, expectation invariant. We provide a self-contained proof of this result and complement it by a number of weak translation invariance conditions and by a dual condition expressed in terms of the conjugate functional.

\begin{theorem}
\label{th: zero collapse convex}
Let $\ph\colon\cX\to(-\infty,\infty]$ be proper, convex, $\sigma(\cX,\cX^\ast)$-lower semicontinuous, and law invariant. Then, the following statements are equivalent:
\begin{enumerate}[(i)]
    \item $\ph$ is expectation invariant.
    \item $\ph$ is the supremum of a family of affine functions of the expectation.
    \item There exists a nonconstant $Z\in\cX$ with $\E[Z]=0$ such that
\[
\ph(X+tZ)=\ph(X), \ \ \ X\in\cX, \ t\in\R.
\]
    \item There exist $a\in\R$ and a nonconstant $Z\in\cX$ with $\E[Z]=0$ such that
\[
\ph(X+tZ)=\ph(X)+at, \ \ \ X\in\cX, \ t\in\R.
\]
    \item For every $X\in\cX$ there exists a nonconstant $Z_X\in\cX$ with $\E[Z_X]=0$ such that
\[
\ph(X+tZ_X)\leq\ph(X), \ \ \ t\geq0.
\]
\item There exist $X\in\dom(\ph)$ and a nonconstant $Z\in\cX$ with $\E[Z]=0$ such that
\[
\ph(X+tZ)\leq\ph(X), \ \ \ t\geq0.
\]
    \item $\dom(\ph^\ast)\subset\R$.
\end{enumerate}
\end{theorem}
\begin{proof}
It is straightforward to verify that (ii) implies (iii), which in turn implies (iv), and that (v) implies (vi). Also note that $\dom(\ph)\cap\R\neq\emptyset$ by dilatation monotonicity recorded in Proposition~\ref{prop: equivalence law invariance}.

(i) implies (ii): If (i) holds, then Proposition \ref{prop: dual representation} yields for every $X\in\cX$
\[
\ph(X) = \ph(\E[X]) = \sup_{Y\in\cX^\ast}\{\E[\E[X]Y]-\ph^\ast(Y)\} = \sup_{Y\in\dom(\ph^\ast)}\{\E[Y]\E[X]-\ph^\ast(Y)\}.
\]
(iv) implies (vii): This is a direct consequence of Proposition \ref{prop: dual representation} and Theorem~\ref{theo:meta}.

(vii) implies (v): This is a direct consequence of Proposition~\ref{prop: dual representation}.

(vi) implies (i): Let $X$ and $Z$ be as in the assertion of (vi) and consider the nonempty convex set $\CC:=\{V\in\CX\,;\ \ph(V)\le\ph(X)\}$. As $Z\in\CC^\infty$, it follows from Proposition~\ref{prop: law invariance recession cone collapse} that $\dom(\sigma_\CC)\subset\R$. Note that, for every $Y\in\dom(\ph^*)$,
\[\sigma_\CC(Y)=\sup_{V\in\CC}\{\E[VY]-\ph(V)+\ph(V)\}\le \ph^*(Y)+\ph(X)<\infty.\]
Hence, $\dom(\ph^*)\subset\R$. Together with Proposition~\ref{prop: dual representation}, for every $V\in\cX$
\[
\ph(V) = \sup_{Y\in\dom(\ph^*)}\{\E[VY]-\ph^\ast(Y)\}=\sup_{Y\in\dom(\ph^*)}\{\E[Y]\E[V]-\ph^\ast(Y)\}=\ph(\E[V]).
\]
This concludes the proof of the equivalence.
\end{proof}

We turn to revisiting \cite[Theorem 4.5]{Collapse}. This result states that, under convexity and $\sigma(\cX,\cX^\ast)$-lower semicontinuity, a functional that is law invariant and translation invariant along a nonconstant random variable with nonzero expectation must collapse to the mean up to an affine transformation. We provide a compact proof of this result and complement it by a dual condition expressed in terms of the conjugate functional.

\begin{theorem}
\label{thm:affine}
Let $\ph\colon\cX\to(-\infty,\infty]$ be proper, convex, $\sigma(\cX,\cX^\ast)$-lower semicontinuous, and law invariant. Then, the following statements are equivalent:
\begin{enumerate}[(i)]
    \item $\ph$ is an affine function of the expectation.
    \item There exist $a\in\R$ and a nonconstant $Z\in\CX$ with $\E[Z]\neq0$ such that
\[
\ph(X+tZ)=\ph(X)+at, \ \ \ X\in\cX, \ t\in\R.
\]
   \item There exist $a\in\R$, a nonconstant $Z\in\CX$ with $\E[Z]\neq0$, and $x\in\dom(\ph)\cap\R$ such that
\[
\ph(x+tZ)=\ph(x)+at, \ \ \ t\in\R.
\]
    \item $\dom(\ph^\ast)\subset\R$ and $\vert\dom(\ph^\ast)\vert=1$.
\end{enumerate}
\end{theorem}
\begin{proof}
It is clear that (i) implies (ii), which in turn implies (iii). Now, assume that (iii) holds.
By Proposition~\ref{prop: dual representation} and Theorem~\ref{theo:meta}, $\emptyset\neq\dom(\ph^*)\subset\R$. Moreover, each $y\in\dom(\ph^*)$ must satisfy
\[
\sup_{t\in\R}\{(y\E[Z]-a)t\}+yx-\ph(x) = \sup_{t\in\R}\{\E[(x+tZ)y]-\ph(x+tZ)\} \leq \ph^\ast(y) < \infty,
\]
showing that $\dom(\ph^*)=\{\tfrac{a}{\E[Z]}\}$. The proof that (iii) implies (iv) is complete. Finally, assume that (iv) holds and let $y\in\R$ be (the unique scalar) such that $\ph^*(y)<\infty$. It immediately follows from Proposition~\ref{prop: dual representation} that
\[
\ph(X) = \E[Xy]-\ph^\ast(y) = y\E[X]-\ph^\ast(y).
\]
This shows that (iv) implies (i) and concludes the proof of the equivalence.
\end{proof}


\subsection{Collapse to the mean: The quasiconvex case}
\label{sect: quasiconvex}

In this section we investigate to which extent the collapse to the mean documented above generalises to quasiconvex functionals. It should be noted that, being heavily based on conjugate duality, the proofs in the convex case do not admit a direct adaptation to the quasiconvex case. In fact, we tackle the collapse to the mean in our more general setting by pursuing a different strategy based on the analysis of recession directions and their interaction with law invariance discussed in Section~\ref{sect: general principle}.

Our first result establishes that Theorem \ref{th: zero collapse convex} continues to hold if we replace convexity with quasiconvexity provided the condition involving conjugate functions is appropriately adapted to a condition involving sublevel sets. In the accompanying remark we show the link between these two conditions.

\begin{theorem}\label{thm:main}
Let $\ph\colon\cX\to(-\infty,\infty]$ be proper, quasiconvex, $\sigma(\cX,\cX^\ast)$-lower semicontinuous, and law invariant. Then, the following statements are equivalent:
\begin{enumerate}[(i)]
    \item $\ph$ is expectation invariant.
    \item There exists a nonconstant $Z\in\cX$ with $\E[Z]=0$ such that
\[
\ph(X+tZ)=\ph(X), \ \ \ X\in\cX, \ t\in\R.
\]
    \item For every $X\in\cX$ there exists a nonconstant $Z_X\in\cX$ with $\E[Z_X]=0$ such that
\[
\ph(X+tZ_X)\leq\ph(X), \ \ \ t\geq0.
\]
    \item For every $m\in\R$ we have $\dom(\sigma_{\{\ph\leq m\}})\subset\R$.
\end{enumerate}
\end{theorem}
\begin{proof}
It is clear that (i) implies (ii), which in turn implies (iii). Now, assume that (iii) holds. Take $m\in\R$ and set $\cC_m=\{\ph\leq m\}$. If $\cC_m=\emptyset$, then we have $\dom(\sigma_{\cC_m})=\emptyset$. Hence, suppose that $\cC_m\neq\emptyset$ and take any $X\in\cC_m$. By assumption, for every $t\geq0$ we have $X+tZ_X\in\cC_m$. This implies that $Z_X\in\cC_m^\infty$. It follows from Proposition~\ref{prop: law invariance recession cone collapse} that $\dom(\sigma_{\cC_m})\subset\R$, showing that (iii) implies (iv). Finally, assume that (iv) holds. For every $m\in\R$ set again $\cC_m=\{\ph\leq m\}$. As $\dom(\sigma_{\cC_m})\subset\R$ and $\sigma_{\cC_m}$ is positively homogeneous, it follows from Proposition \ref{prop: dual representation} that
\[
\cC_m = \{X\in\cX \,; \ -\sigma_{\cC_m}(-1)\leq\E[X]\leq\sigma_{\cC_m}(1)\}.
\]
As a consequence, we obtain for every $X\in\cX$
\[
\ph(X) = \inf\{m\in\R \,; \ X\in\cC_m\} = \inf\{m\in\R \,; \ -\sigma_{\cC_m}(-1)\leq\E[X]\leq\sigma_{\cC_m}(1)\}.
\]
In particular, $\ph(X)=\ph(\E[X])$ for every $X\in\cX$. This shows that (iv) implies (i).
\end{proof}

\begin{remark}
Let $\ph\colon\cX\to(-\infty,\infty]$ be proper, convex, and $\sigma(\cX,\cX^\ast)$-lower semicontinuous. Moreover, take $m\in\R$ such that $\{\ph\leq m\}\neq\emptyset$. It was shown in the proof of Theorem \ref{th: zero collapse convex} that $\dom(\ph^\ast)\subset\R$ (point (vii) in Theorem \ref{th: zero collapse convex}) is a direct consequence of $\dom(\sigma_{\{\ph\leq m\}})\subset\R$ (point (iv) in Theorem \ref{thm:main}).
\end{remark}

The next example shows that point (vi) in Theorem \ref{th: zero collapse convex} is specific to the convex case and cannot be added to the equivalent conditions in Theorem \ref{thm:main}.

\begin{example}
\label{ex: key example}
Let the functional $\rho\colon \CX\to\R$ be defined by
\[
\rho(X)=\frac{1}{2}\E[X]+\int_{1/2}^1q_X(s)ds.
\]
Note that $\rho$ is convex, $\sigma(\CX,L^\infty)$-lower semicontinuous, and law invariant. Set
\[
\cC_m =
\begin{cases}
\{X\in \CX \,; \ \rho(X)\le m\} & \mbox{if} \ m<0,\\
\{X\in \CX \,; \ \E[X]\le2m\} & \mbox{if} \ m\ge 0.
\end{cases}
\]
Define the functional $\ph\colon\CX\to\R$ by setting
\[
\ph(X)=\inf\{m\in\R \,; \ X\in\cC_m\}=
\begin{cases}
\rho(X) & \mbox{if} \ \rho(X)<0,\\
\frac{1}{2}\max\{\E[X],0\} & \mbox{if} \ \rho(X)\geq0.
\end{cases}
\]
For all $X\in\CX$ and $m\in\R$ we have $\ph(X)\leq m$ if and only if $X\in\cC_m$, showing that $\ph$ is quasiconvex and $\sigma(\CX,L^\infty)$-lower semicontinuous. Moreover, $\ph$ is clearly law invariant and satisfies $\ph(0)=0$. Now, use nonatomicity to find a random variable $Z$ such that
\[
\P(Z=2)=1-\P(Z=-1)=\tfrac 1 3.
\]
A direct calculation shows that $\E[Z]=0$ and $\rho(Z)=\frac{1}{2}$. As a result, we obtain for every $m\geq0$ that $\ph(0+mZ) = m\ph(Z) = 0 = \ph(0)$, showing that $\ph$ satisfies point (vi) in Theorem \ref{th: zero collapse convex}. However, $\ph$ is not expectation invariant. To see this, compare a random variable $X$ with $\P(X=4)=\P(X=-6)=\tfrac 1 2$ to the constant random variable $Y=-1$. Then, we have $\E[X]=\E[Y]=-1$, but $\rho(X)=0=\ph(X)$, while $\rho(Y)=-1=\ph(Y)$.
\end{example}

We turn to the collapse to the mean established in Theorem \ref{thm:affine}. The next result shows that, if convexity is relaxed to quasiconvexity, then the collapse to the mean continues to hold in the presence of translation invariance (point (ii) in Theorem \ref{thm:affine}).

\begin{theorem}\label{thm:extquasiconv}
Let $\ph\colon\cX\to(-\infty,\infty]$ be proper, quasiconvex, $\sigma(\cX,\cX^\ast)$-lower semicontinuous, and law invariant. Then, the following statements are equivalent:
\begin{enumerate}[(i)]
    \item $\ph$ is an affine function of the expectation.
    \item There exist $a\in\R$ and a nonconstant $Z\in\CX$ with $\E[Z]\neq0$ such that
\[
\ph(X+tZ)=\ph(X)+at, \ \ \ X\in\cX, \ t\in\R.
\]
\item There exist $a\in\R$ and a nonconstant $Z\in\CX$ with $\E[Z]\neq0$ such that
\[
\ph(x+tZ)=\ph(x)+at, \ \ \ x\in\R, \ t\in\R.
\]
\end{enumerate}
\end{theorem}
\begin{proof}
It is easy to see that (i) implies (ii) and that (ii) implies (iii). Assume now that (iii) holds. Suppose $m\in\R$ is such that $\{\ph\le m\}\neq\emptyset$. By dilatation monotonicity of $\ph$ recorded in Proposition \ref{prop: equivalence law invariance}, we find $x\in\R$ such that $\ph(x)\le m$. Making use of dilatation monotonicity once more, we infer for all $t\ge 0$ that
\[
\ph\left(x+t(\E[Z]-Z)\right)=\ph(x+t\E[Z])-at\le \ph(x+tZ)-at=\ph(x)\le m.
\]
As $U:=\E[Z]-Z$ belongs to the recession cone of $\{\ph\leq m\}$ and $\E[U]=0$, Proposition~\ref{prop: law invariance recession cone collapse} implies that $\dom(\sigma_{\{\ph\le m\}})\subset\R$. By Theorem~\ref{thm:main}, $\ph$ is expectation invariant. In particular,
\begin{align*}\ph(X)&	=\ph\left(X-\tfrac{\E[X]}{\E[Z]}Z+\tfrac{\E[X]}{\E[Z]}Z\right)=\ph\left(X-\tfrac{\E[X]}{\E[Z]}Z\right)+\tfrac{a}{\E[Z]}\E[X]\\
&=\ph\left(\E\left[X-\tfrac{\E[X]}{\E[Z]}Z\right]\right)+\tfrac{a}{\E[Z]}\E[X]=\ph(0)+\tfrac{a}{\E[Z]}\E[X]
\end{align*}
for every $X\in\CX$. That is, $\ph$ is an affine function of the expectation as stated in (i).
\end{proof}

The following example shows that point (iii) in Theorem~\ref{thm:affine} fails to produce a collapse to the mean under mere quasiconvexity. In particular, this observation holds no matter the value of the expectation of the nonconstant random variable along which local translation invariance in the sense of point (iii) in Theorem~\ref{thm:affine} holds. Moreover, the example demonstrates that Theorem~\ref{thm:extquasiconv} cannot be improved.

\begin{example}\label{ex:quasiconv}
Consider the setting of Example \ref{ex: key example}, and let the random variable $Z$ be as described there, i.e., $\P(Z=2)=1-\P(Z=-1)=\tfrac 13$. Moreover, let $U$ be a random variable whose distribution is given by $\P(U=4)=\P(U=0)=\tfrac 1 2$. Both $Z$ and $U$ are nonconstant, $\E[Z]=0$, and $\E[U]=2$. We have already observed that $\ph(tZ)=0$, $t\ge 0$. One can also verify that $\rho(-tZ)=\tfrac t 2\ge 0$ which means that $\ph(-tZ)=0$, $t\ge 0$.
Moreover, for every $t\ge 0$,
\[\ph(-tU)=t\rho(-U)=-t,\quad \ph(tU)=\tfrac t 2\E[U]=t.\]
Hence, $\ph$ satisfies point (iii) in Theorem \ref{thm:affine} even without the additional condition on the expectation. However, $\ph$ is neither an affine nor a convex function of the expectation (and not even expectation invariant, as observed in Example~\ref{ex: key example}).
\end{example}


\subsection{Collapse to the mean: The case of consistent risk measures}
\label{sect: consistent risk measures}

In this and the following section, we establish a collapse to the mean for classes of law-invariant functionals beyond the quasiconvex family. In this section, we focus on functionals that are translation invariant along constants and monotonic with respect to second-order stochastic dominance. Following the terminology in~\cite{Consistent}, we refer to them as consistent risk measures. This class covers the family of law-invariant convex risk measures but also includes nonconvex functionals, e.g., minima of law-invariant convex risk measures. As translation invariance along constants implies that convexity and quasiconvexity are equivalent, the class of consistent risk measures contains functionals that are not quasiconvex. As a result, we cannot resort to the quasiconvex results in Section~\ref{sect: quasiconvex}.

First, recall that a \emph{consistent risk measure} is a proper functional $\ph:\CX\to(-\infty,\infty]$ that is:
\begin{enumerate}
\item[(1)] \emph{cash-additive}, i.e., $\ph(X+m)=\ph(X)+m$ for all $X\in\CX$ and $m\in\R$.
\item[(2)] {\em consistent with second-order stochastic dominance}, i.e., for all $X,Y\in\CX$,
\[
\mbox{$\E[f(X)]\geq\E[f(Y)]$ for every nondecreasing convex $f\colon\R\to\R$} \ \implies \ \ph(X)\geq\ph(Y).
\]
\item[(3)] {\em normalised}, i.e., $\ph(0)=0$.\footnote{~In \cite{Consistent} only the condition $\ph(0)\in\R$ is imposed on a consistent risk measure $\ph\colon L^\infty\to\R$. Applying an affine transformation to $\ph$, one can guarantee $\ph(0)=0$ though.}
\end{enumerate}

Given its defining properties, a consistent risk measure takes only finite values on $L^\infty$. Moreover, every consistent risk measure is automatically dilatation monotone and law invariant by property (2).
In case $\CX=L^\infty$, every normalised, law-invariant, and convex risk measure is a consistent risk measure. The same holds for normalised, law-invariant, $\sigma(\CX,\CX^*)$-lower semicontinuous convex risk measures by Proposition~\ref{prop: equivalence law invariance}. The next proposition shows that every consistent risk measure on $L^\infty$ can be extended uniquely to a $\sigma(\CX,\CX^*)$-lower semicontinuous consistent risk measure. In particular, a consistent risk measure on $L^\infty$ is automatically $\sigma(L^\infty,\cX^*)$-lower semicontinuous.

\begin{proposition}
Let $\ph\colon L^\infty\to\R$ be a consistent risk measure. Then, there is a unique, $\sigma(\CX,\CX^*)$-lower semicontinuous, consistent risk measure $\overline\ph\colon\CX\to(-\infty,\infty]$ that extends $\ph$.
\end{proposition}
\begin{proof}
Note that $\ph$ is dilatation monotone in the sense of \cite{Xanthos}. In addition, by \cite[Theorem 3.5]{Consistent}, $\ph$ has the Fatou property, i.e., for every uniformly bounded sequence $(X_n)\subset L^\infty$ converging to $X\in L^\infty$ almost surely, $\ph(X)\le\liminf_{n\to\infty}\ph(X_n)$. Let $\Pi$ denote the set of finite measurable partitions of $\Omega$. For $X\in L^1$ and $\pi\in\Pi$ we write $\E[X|\pi]:=\E[X|\sigma(\pi)]$, where $\sigma(\pi)$ is the $\sigma$-field generated by $\pi$. \cite[Theorem 4]{Xanthos} proves that the functional $\ph^\sharp\colon L^1\to(-\infty,\infty]$ defined by
\[\ph^\sharp(X):=\sup_{\pi\in\Pi}\ph(\E[X|\pi]),\]
is a $\sigma(L^1,L^\infty)$-lower semicontinuous, dilatation monotone in the sense of \cite{Xanthos}, cash-additive extension of $\ph$. \textit{A fortiori}, the restriction of $\ph^\sharp$ to $\cX$, denoted by $\overline\ph$, is a $\sigma(\CX,\CX^*)$-lower semicontinuous, dilatation monotone in the sense of \cite{Xanthos}, cash-additive extension of $\ph$. It remains to verify consistency of $\ph^\sharp$, which implies that of $\overline\ph$. By \cite[Theorem B.3]{Consistent}, it suffices to check for dilatation monotonicity in the sense of \cite{Consistent}. To this end, suppose $X,Y\in L^1$ satisfy $\E[Y|X]=X$. Let $(\pi_n)\subset\sigma(\CX)$ be an increasing sequence of finite measurable partitions such that $X_n=\E[X|\pi_n]\to X$ in $L^1$. For all $n\in\N$, $\E[Y|X_n]=\E[X|\pi_n]$ holds, which entails
\[\ph^\sharp(Y)\ge\limsup_{n\to\infty}\ph^\sharp(X_n)\ge\liminf_{n\to\infty}\ph^\sharp\left(\E[X|\pi_n]\right)\ge\ph^\sharp(X)=\ph^\sharp(\E[Y|X]).\]
This is the desired dilatation monotonicity of $\ph^\sharp$.
Uniqueness of $\overline\ph$ can be seen to be a consequence of the uniqueness statement in \cite[Theorem 4]{Xanthos}.
\end{proof}

The following representation result from \cite{Consistent} will play a crucial role in our later analysis. In the terminology of \cite{adjustedES}, it shows that any consistent risk measure on $L^\infty$ can be expressed as a minimum of adjusted Expected Shortfalls.

\begin{lemma}[{\cite[Theorem 3.1]{Consistent}}]
\label{lem: ES representation}
The \tn{Expected Shortfall} of $X\in\CX$ at level $p\in[0,1]$ is
\[
\ES_p(X):=
\begin{cases}
\frac 1 {1-p}\int_{p}^1q_X(s)ds & \mbox{if} \ p<1,\\
\inf\{x\in\R \,; \ \P(X\le x)=1\} & \mbox{if} \ p=1.
\end{cases}
\]
Let $\ph\colon L^\infty\to\R$ be a consistent risk measure. Then, for every $X\in L^\infty$,
\[
\ph(X) = \min_{Y\in \CA_\ph}\sup_{p\in[0,1]}\{\ES_p(X)-\ES_p(Y)\}.
\]
where
\[\CA_\ph:=\{Y\in L^\infty\,;\ \ph(Y)\le 0\}\]
denotes the \tn{acceptance set} of $\ph$.
\end{lemma}

Our main result establishes a collapse to the mean for consistent risk measures. We show that linearity along a nonconstant random variable is sufficient to reduce the functional to a standard expectation. In line with our previous result, we also provide an equivalent condition for the collapse in terms of directions of recession and conjugate functions.

\begin{theorem}
\label{thm:consistent}
Let $\ph\colon\CX\to(-\infty,\infty]$ be a $\sigma(\CX,\CX^*)$-lower semicontinuous consistent risk measure. Then, the following are equivalent:
\begin{enumerate}[(i)]
\item $\ph$ coincides with the expectation under $\probp$.
\item There exist a nonconstant $Z\in \CX$ and $a\in\R$ such that
\[\ph(tZ)=at,\quad t\in\R.\]
\item There exists a nonconstant $U\in\CX$ such that $\E[U]=0$ and
\[\sup_{t\ge 0}\ph(tU)\le 0.\]
\end{enumerate}
Any of the previous statements implies:
\begin{enumerate}[(iv)]
\item $\dom(\ph^*)=\{1\}$.
\end{enumerate}
Statements (i)--(iv) are equivalent if, additionally,
\begin{equation}\label{eq:star}\ph(\lambda X)\leq\lambda\ph(X),\quad X\in \CX,\,\lambda\in[0,1].\end{equation}
\end{theorem}
\begin{proof}
It is trivial to see that (i) implies (ii). In order to see that (ii) implies (iii), recall first that $\ph$ is dilatation monotone as observed above. Hence, we may estimate
\[a=\ph(Z)\ge\ph(\E[Z])=\E[Z]=-\E[-Z]=-\ph(\E[-Z])\ge -\ph(-Z)=a.\]
This means that $a=\E[Z]$. Set $U=Z-\E[Z]$ and use cash-additivity of $\ph$ to infer for every $t\ge 0$ that
$$\ph(tU)=\ph(tZ-t\E[Z])=\ph(tZ)-t\E[Z]=ta-t\E[Z]=0.$$
This yields the desired implication.

Now, we claim that (iii) implies (i). We first consider the case $\CX=L^\infty$ and fix an arbitrary $X\in L^\infty$. Using Lemma~\ref{lem: ES representation}, we have
\begin{equation}
\label{eq: consistent 1}
\ph(X)\le\inf_{t>0}\sup_{p\in[0,1]}\{\ES_p(X)-t\ES_p(U)\}.
\end{equation}
As $\E[U]=0$ by assumption, Lemma~\ref{lem:interior} implies that $\ES_p(U)>0$ for every $p\in(0,1)$. Let $q\in(0,1)$ be arbitrary and choose $t_0>0$ such that $\ES_1(X)-t_0\ES_q(U)\le\E[X]$. Note that
\[
\inf_{t>0}\sup_{p\in[0,1]}\{\ES_p(X)-t\ES_p(U)\} = \inf_{t>t_0}\sup_{p\in[0,1]}\{\ES_p(X)-t\ES_p(U)\}.
\]
Moreover, for all $p\in[q,1]$ and $t>t_0$,
\[\ES_p(X)-t\ES_p(U)\le \ES_1(X)-t_0\ES_q(U) \le \E[X] = \ES_0(X)-t\ES_0(U).\]
As a result, we get
\begin{equation}
\label{eq: consistent 2}
\inf_{t>0}\sup_{p\in[0,1]}\ES_p(X)-t\ES_p(U)=\inf_{t>t_0}\sup_{p\in[0,q]}\ES_p(X)-t\ES_p(U).
\end{equation}
Now, for all $p\in[0,q]$,
\[|\ES_p(X)-\E[X]|=-\tfrac 1{1-p}\int_0^pq_X(s)ds+\tfrac{p}{1-p}\int_0^1q_X(s)ds\le\tfrac{2q}{1-q}\|X\|_\infty.\]
Combining this inequality with \eqref{eq: consistent 1} and \eqref{eq: consistent 2} yields
\[\ph(X)\le \inf_{t>t_0}\left\{\E[X]+\tfrac{2q}{1-q}\|X\|_\infty-\inf_{p\in[0,q]}t\ES_p(U)\right\}=\E[X]+\tfrac{2q}{1-q}\|X\|_\infty.\]
We conclude by noting that, by dilatation monotonicity,
\[\E[X]=\ph(\E[X])\le \ph(X)\le\lim_{q\downarrow 0}\{\E[X]+\tfrac{2q}{1-q}\|X\|_\infty\}=\E[X].\]
This shows that $\ph(X)=\E[X]$ whenever $X\in L^\infty$. To conclude the proof of the implication, we consider the case of a general space $\CX$. Note that for an arbitrary finite sub-$\sigma$-algebra such that $\E[U|\CG]\in L^\infty$ is nonconstant, dilatation monotonicity implies
\[\sup_{t\ge 0}\ph\left(t\E[U|\CG]\right)=\sup_{t\ge 0}\ph(tU)\le 0.\]
The preceding argument shows that $\ph$ coincides with the expectation under $\probp$ when restricted to $L^\infty$. By, e.g., \cite[Lemma 4.1]{Bellini}, $L^\infty$ is dense in $\CX$ with respect to $\sigma(\CX,\CX^\ast)$. Take a net $(X_\alpha)\subset L^\infty$ satisfying $X_\alpha\to X$ with respect to $\sigma(\CX,\CX^\ast)$. By dilatation monotonicity and $\sigma(\CX,\CX^\ast)$-lower semicontinuity,
\[
\E[X]=\ph(\E[X])\leq \ph(X) \leq \liminf_\alpha\ph(X_\alpha) = \liminf_\alpha\E[X_\alpha] = \E[X].
\]
This delivers (i). Clearly, (i) implies (iv). We conclude by proving that (iv) implies (iii) under the additional assumption that $\ph(\lambda X)\leq\lambda\ph(X)$ for all $\lambda\in[0,1]$ and $X\in\CX$. To this end, let $A\in\CF$ satisfy $\P(A)=\tfrac 1 2$ and set $\CG=\{\emptyset,A,A^c,\Omega\}$. For every $\CG$-measurable, positive, nonconstant $Y\in L^\infty$ with $\E[Y]=1$ and for every $n\in\N$ we claim that
\begin{equation}\label{infinite}
\sup\{\E[XY] \,; \ X\in\CA_\ph, \ \mbox{$X$ is $\CG$-measurable}, \ \|X\|_\infty>n\}=\infty.
\end{equation}
To see this, observe that
\[
\sup\{\E[XY] \,; \ X\in\CA_\ph, \ \mbox{$X$ is $\CG$-measurable}, \ \|X\|_\infty\leq n\} \leq n\E[Y] < \infty.
\]
At the same time,
\[
\sup\{\E[XY] \,; \ X\in\CA_\ph, \ \mbox{$X$ is $\CG$-measurable}\} = \sup_{X\in\CA_\ph}\E[\E[X|\CG]Y],
\]
where we used that $\E[X|\CG]\in\CA_\ph$ holds for every $X\in\CA_\ph$ by dilatation monotonicity. As a consequence, by $\CG$-measurability of $Y$,
\[
\sup\{\E[XY] \,; \ X\in\CA_\ph, \ \mbox{$X$ is $\CG$-measurable}\} = \sup_{X\in\CA_\ph}\E[XY] = \ph^\ast(Y) = \infty.
\]
This delivers~\eqref{infinite}. Now, for $n\in\N$ define $Y_n=\tfrac{n-1}n\ind_A+\tfrac{n+1}n\ind_{A^c}\in L^\infty$ and note that $Y_n$ is $\CG$-measurable, positive, nonconstant, and satisfies $\E[Y_n]=1$. It follows from~\eqref{infinite} that we find a $\CG$-measurable $X_n\in\{ \ph\leq0\}$ with $\|X_n\|_\infty>n$ and $\E[X_nY_n]\ge 1$. As $\E[X_n]\E[Y_n]=\E[X_n]\le\ph(X_n)\le0$ by dilatation monotonicity and cash-additivity, $X_n$ cannot be constant by Lemma~\ref{lem:interior}.
Using compactness of the appropriate unit sphere in $\R^2$, we can assume without loss of generality that there is a suitable $\CG$-measurable $U\in L^\infty$ such that $U\neq0$ and
\[
\frac{X_n}{\|X_n\|_\infty}\to U.
\]
By our additional assumption, for every $t>0$ we eventually have $t\frac{X_n}{\|X_n\|_\infty}\in\CA_\ph$ and, thus, $tU\in\CA_\ph$ or, equivalently, $\ph(tU)\leq0$. To prove (iii), it remains to show that $\E[U]=0$. To this effect, note that  $\frac{X_nY_n}{\|X_n\|_\infty}\to U$. As a result, applying dilatation monotonicity again,
\[
0\geq \ph(U) \geq \E[U] = \lim_{n\to\infty}\frac{\E[X_nY_n]}{\|X_n\|_\infty} \geq \lim_{n\to\infty}\frac{1}{\|X_n\|_\infty} = 0.
\]
This concludes the proof.
\end{proof}

\begin{remark}
Condition \eqref{eq:star} means that the risk measure $\ph$ is {\em star shaped} in the sense of \cite{Starshape}. By \cite[Proposition 2]{Starshape}, the latter is equivalently characterised by the fact that the acceptance set $\CA_\ph$ is star shaped about 0. Consistent risk measures satisfying \eqref{eq:star} are characterised in \cite[Theorem 11]{Starshape}, but we would like to motivate here that, in fact, \eqref{eq:star} is a very mild constraint.
By \cite[Theorem 3.3]{Consistent} or Lemma~\ref{lem: ES representation} above, a consistent risk measure $\ph\colon L^\infty\to\R$ is represented by a family  $\mathcal T$ of convex law-invariant risk measures $\tau$ in that
\[\ph(X)=\inf_{\tau\in\mathcal T}\tau(X),\quad X\in L^\infty.\]
If each $\tau\in\mathcal T$ is normalised, i.e., $\tau(0)=0$, then $\ph$ has property \eqref{eq:star}.
\end{remark}


\subsection{Collapse to the mean: The case of Choquet integrals}
\label{sect: Choquet}

As mentioned in the introduction, the research on law-invariant functionals and their collapse to the mean was triggered by \cite{Castagnoli}, where the focus was on Choquet integrals associated with special submodular law-invariant capacities. The property of submodularity is equivalent to convexity of the Choquet integral. As such, the collapse to the mean established there can be seen as a special case of the results in Section~\ref{sect: convex}. In this section, we extend the collapse to the mean to nonconvex Choquet integrals. To this effect, it should be noted that we cannot resort to the quasiconvex results in Section~\ref{sect: quasiconvex} because, for a Choquet integral, quasiconvexity automatically implies convexity in view of translation invariance along constants.

We start by recalling some basic notions. A {\em capacity} is a function $\mu\colon\CF\to[0,1]$ such that $\mu(\emptyset)=0$ and $\mu(\Omega)=1$, and $\mu(A)\leq\mu(B)$ for all $A,B\in\CF$ with $A\subset B$. We say that $\mu$ is:
\begin{enumerate}[(1)]
    \item {\em coherent} if there exists a family $\CQ$ of probability measures $\probq\colon\CF\to[0,1]$ such that
\[
\mu(A)=\sup_{\probq\in\CQ}\probq(A), \ \ \ A\in\CF.\footnote{~Coherent capacities are also known in the literature as {\em upper probabilities} or {\em upper envelopes}.}
\]
    \item {\em submodular} if, for all $A,B\in\CF$,
\[
\mu(A\cup B)+\mu(A\cap B) \leq \mu(A)+\mu(B).\footnote{~Submodular capacities are sometimes called {\em 2-alternating}.}
\]
    \item {\em law invariant} if, for all $A,B\in\CF$,
\[
\probp(A)=\probp(B) \ \implies \ \mu(A)=\mu(B).\footnote{~Law-invariant capacities are sometimes called {\em symmetric}.}
\]
\end{enumerate}
We also recall that the {\em dual capacity} $\overline\mu\colon\CF\to[0,1]$ is defined by
\[\overline\mu(A):=1-\mu(A^c).\]
In what follows, we denote by $\CL^\infty$ the space of bounded measurable functions $X\colon\Omega\to\R$. The {\em Choquet integral} associated with a capacity $\mu$ is the functional $\E_\mu\colon\CL^\infty\to\R$ defined by
\[
\E_\mu[X] := \int_{-\infty}^0(\mu(X>s)-1)ds+\int_0^\infty\mu(X>s)ds.
\]
If $\mu$ is countably additive, i.e., a probability measure, then the Choquet integral reduces to a standard expectation. The next proposition collects some well-known facts about Choquet integrals. In particular, note that, under a law-invariant capacity, we can unambiguously define the Choquet integral on the space $L^\infty$ as will be tacitly done below.

\begin{proposition}\label{prop:Choquet}
Let $\mu$ be a capacity. Then, the following statements hold:
\begin{enumerate}[(i)]
   \item For every $X\in \CL^\infty$ we have $\E_\mu[X]=-\E_{\overline\mu}[-X]$.
   \item For all $X\in \CL^\infty$, $t\geq0$, and $c\in\R$, we have $\E_\mu[tX+c]=t\E_\mu[X]+c$.
   \item $\E_\mu$ is convex if and only if $\mu$ is submodular.
   \item $\E_\mu$ is law invariant if and only if $\mu$ is law invariant.
\end{enumerate}
\end{proposition}

By the classical results in \cite{Schmeidler}, a submodular capacity is automatically coherent. The converse does not hold in general; see, e.g., \cite{Wassermann}. We target the extension of Theorem \ref{thm:affine} to nonconvex Choquet integrals associated with coherent capacities. In fact, we shall go one step further and focus on so-called Jaffray-Philippe (JP) capacities introduced in \cite{JP}. A capacity $\mu$ is a {\em JP capacity} if there is a pair $(\nu,\alpha)$ of a coherent capacity $\nu$ and $\alpha\in[0,1]$ such that\footnote{~It has already been observed in \cite{JP} that the case $\alpha=\tfrac 1 2$ is peculiar, hence we exclude it from our results.}
\[
\mu(A)=\alpha\nu(A)+(1-\alpha)\overline\nu(A), \ \ \ A\in\CF,
\]
where $\overline\nu$ is the dual capacity of $\nu$.
JP capacities encompass both submodular and coherent capacities, as well as neo-additive capacities introduced in \cite{Chateauneuf}.\footnote{~For a probability measure $\Q$ and $\delta,\alpha\in[0,1]$, the {\em neo-additive} capacity defined by
\[
\mu(A)=(1-\delta)\Q(A)+(1-\alpha)\delta\ind_{\CF\setminus\{\emptyset\}}(A)+\alpha\delta\ind_{\{\Omega\}}(A), \ \ \ A\in\CF,
\]
is the JP-capacity generated by $(\nu,1-\alpha)$, where $\nu=(1-\delta)\Q+\delta\ind_{\CF\setminus\{\emptyset\}}$ is a submodular capacity.} A first lemma characterises law invariance of JP capacities.

\begin{lemma}
\label{lem: law invariance JP}
Let $\mu$ be a JP capacity represented by the pair $(\nu,\alpha)$, where $\alpha\neq \tfrac 1 2$. Then, $\mu$ is law invariant if and only if $\nu$ is law invariant.
\end{lemma}
\begin{proof}
Law invariance of the capacity $\nu$ implies law invariance of the dual capacity $\overline\nu$ and thus of $\mu$. Conversely, assume that $\mu$ is law invariant. Its dual capacity is given by $\overline\mu=\alpha\overline\nu+(1-\alpha)\nu$.
As $\alpha\neq \tfrac 1 2$, we may recover $\nu$ as
\begin{equation}\label{eq:polarisation}\nu = \tfrac{\alpha}{2\alpha-1}\mu-\tfrac {1-\alpha}{2\alpha-1}\overline\mu.
\end{equation}
As the dual capacity $\overline\mu$ is also law invariant, the value of the right-hand side in \eqref{eq:polarisation} only depends on the $\P$-probability of its argument. This implies law invariance of $\nu$.
\end{proof}

We establish the desired collapse to the mean for nonconvex Choquet integrals. Our result encompasses \cite[Theorem 3.1]{Castagnoli}, which was established under the assumption of submodularity by means of convex duality. Our proof is direct and solely based on Theorem~\ref{theo:meta}.

\begin{theorem}\label{thm:Choquet}
Let $\mu$ be a law-invariant JP capacity represented by a pair $(\nu,\alpha)$. Moreover, assume $\alpha\neq \tfrac 1 2$. Then, the following statements are equivalent:
\begin{enumerate}[(i)]
    \item $\E_\mu$ coincides with the expectation under $\probp$.
    \item There exist $a\in\R$ and a nonconstant $Z\in L^\infty$ such that
\[
\E_\mu[X+tZ]=\E_\mu[X]+at, \ \ \ X\in L^\infty, \ t\in\R.
\]
    \item There exist $a\in\R$ and a nonconstant $Z\in L^\infty$ such that
\[
\E_\mu[tZ]=at, \ \ \ t\in\R.
\]
    \item There exists a nonconstant $Z\in L^\infty$ such that
\[
\E_\mu[-Z]=-\E_\mu[Z].
\]
\end{enumerate}
\end{theorem}
\begin{proof}
Clearly, (i) implies (ii) and (iii) implies (iv). As $\E_\mu[0]=0$, we also see that (ii) implies (iii). Now, suppose that (iv) holds. By point (i) in Proposition~\ref{prop:Choquet}, the assumption reads as $\E_\mu[-Z]=\E_{\overline\mu}[-Z]$ or, equivalently, $\E_\mu[Z]=\E_{\overline\mu}[Z]$. By the polarisation identity in \eqref{eq:polarisation},
\begin{align*}
\E_\nu[-Z]&=\tfrac{\alpha}{2\alpha-1}\E_\mu[-Z]-\tfrac{1-\alpha}{2\alpha-1}\E_{\overline\mu}[-Z]
=\E_\mu[-Z] \\
&=-\E_\mu[Z]
=\tfrac{1-\alpha}{2\alpha-1}\E_{\overline\mu}[Z]-\tfrac{\alpha}{2\alpha-1}\E_\mu[Z]
=-\E_\nu[Z].
\end{align*}
Using point (ii) in Proposition \ref{prop:Choquet}, we conclude that $\E_\nu[tZ]=t\E_\nu[Z]$ for every $t\in\R$. Now, note that $\nu$ is law invariant by Lemma \ref{lem: law invariance JP}. By coherence and the Radon-Nikod\'ym theorem, there exists a family $\mathcal D\subset L^1$ of probability densities such that, for every $A\in\CF$,
\[\nu(A)=\sup_{D\in\mathcal D}\E[D\ind_A].\]
Note furthermore that each $X\in L^\infty$ and each $D\in\mathcal D$ satisfy $\E_\nu[X]\geq\E[DX]$. By Theorem~\ref{theo:meta}, $D$ must be constant. This forces $\nu=\overline\nu=\P$, and consequently $\mu=\probp$, that is, (i) holds. The proof of the equivalence is complete.
\end{proof}


\subsection{Collapse to the mean in optimisation problems}
\label{sect: Optimisation}

In this section we focus on a class of optimisation problems involving law invariance at the level of both the objective function and the optimisation domain.
We investigate the existence of optimal solutions that are antimonotone with respect to a ``pricing density'' appearing in the budget constraint under a list of suitable assumptions.
We prove sharpness of our existence result in the sense that, if any of the listed assumptions is removed, then the result continues to hold only in the trivial situation where the budget constraint ``collapses to the mean''.
This is relevant in applications because a key monotonicity assumption on the optimisation domain is sometimes omitted in the literature, in which case, contrary to what is sometimes stated, the general result cannot be invoked and one has to proceed case by case.

Throughout the entire section we focus on the optimisation problem
\begin{equation*}
\begin{cases}
\ph(X) = \max \\
X\in\cC \\
\E[DX]=p
\end{cases}
\end{equation*}
under the following basic assumptions:
\begin{enumerate}[(1)]
    \item $\ph\colon\cX\to[-\infty,\infty]$ is law invariant,
    \item $\cC\subset\cX$ is law invariant,
    \item $D\in\cX^\ast$ satisfies $\E[D]>0$ and $p\in\R$.
\end{enumerate}
The last constraint is typically interpreted as a budget constraint where $D$ plays the role of a ``pricing density''. We say that the quadruple $(\ph,\cC,D,p)$ is {\em feasible} if the optimisation problem admits an optimal solution. In this case, we denote by $\Max(\ph,\CC,D,p)$ the corresponding optimal value. This problem has been extensively studied in the literature, see, e.g., \cite{Burgert,Carlier,HeZhou,Schied,Xu,Xu2}, and the recent overview in \cite{RueschendorfVanduffel}. In this literature, one encounters the following two types of statements about optimal solutions:
\begin{itemize}
\item There exists an optimal solution that is antimonotone with $D$.
\item All optimal solutions are antimonotone with $D$.
\end{itemize}
As mentioned in the introduction, these statements are very useful because they allow to reduce the original problem to a deterministic optimisation problem involving quantile functions; see, e.g., \cite{RueschendorfVanduffel}.

We start by providing a slight extension to the extant results about existence of optimal solutions that are antimonotone with the ``pricing density''. To this effect, it is convenient to define the following notions:
\begin{enumerate}[(1)]
\item $\CC$ is {\em increasing} if $X+m\in\cC$ for all $X\in\cC$ and $m\ge 0$.
\item $\ph$ is {\em weakly increasing} if $\ph(X+m)\ge\ph(X)$ for all $X\in\CX$ and $m\ge 0$.
\item $\ph$ is {\em increasing} if $\ph(X+m)>\ph(X)$ for all $X\in\CX$ with $\ph(X)\in\R$ and $m>0$.
\end{enumerate}

The next result shows that antimonotone optimal solutions always exist provided that both $\cC$ is increasing and $\ph$ is weakly increasing. If $\ph$ is also increasing, then every optimal solution must be antimonotone with the ``pricing density''.

\begin{theorem}
\label{theo: optimisation}
Let $(\ph,\cC,D,p)$ be a feasible quadruple.
\begin{enumerate}[(i)]
\item If $\CC$ is increasing and $\ph$ is weakly increasing, then there exists an optimal solution that is antimonotone with $D$.
\item If $\CC$ is increasing, $\ph$ is increasing, and $\Max(\ph,\CC,D,p)\in\R$, then all optimal solutions are antimonotone with $D$.
\end{enumerate}
\end{theorem}
\begin{proof}
Let $X\in\cX$ be an optimal solution. To prove (i), let $X'\sim X$ be antimonotone with $D$. Note that $\E[DX']\le\E[DX]$ by Lemma~\ref{lem:interior} and set
\[
m=\frac{\E[DX]-\E[DX']}{\E[D]}\ge0.
\]
As $X\in\cC$, we have $X'\in\cC$ by law invariance of $\cC$. As $\CC$ is increasing, $X'+m\in\cC$. Note that $\E[D(X'+m)]=\E[DX]=p$. In addition, $\ph(X'+m)\ge\ph(X')=\ph(X)$ because the function $\ph$ is weakly increasing and law invariant. We conclude that $X'+m$ is an optimal solution. It remains to observe that $X'+m$ is antimonotone with $D$ by construction.

To establish (ii), assume towards a contradiction that $X$ is not antimonotone with $D$---which entails in particular that $D$ and $X$ are nonconstant---and take $X'$ and $m$ as above. The same argument shows that $X'+m$ is an optimal solution. From Lemma \ref{lem:interior} we derive $m>0$. This yields $\ph(X'+m)>\ph(X')=\ph(X)$ because $\ph$ is increasing and law invariant, and because $\ph(X)\in\R$. However, this contradicts the optimality of $X$. In conclusion, $X$ and $D$ have to be antimonotone.
\end{proof}

The previous result is sometimes stated without the monotonicity assumption on the domain $\cC$ (see, e.g., \cite{RueschendorfVanduffel}) or it is said that the monotonicity assumption on $\cC$ is made without loss of generality (see, e.g., \cite{Xu}).\footnote{~We highlight that the result is also typically stated without the finiteness assumption of the optimal value. This is often justified because the special choice of $\ph$ and $\cC$ ensures finiteness.}
The remainder of the section is devoted to showing that all the assumptions in Theorem \ref{theo: optimisation}, including the monotonicity assumption on $\cC$, are necessary for the result to hold.
More precisely, we show that, if any of the assumptions is removed, then for every choice of a nonconstant ``pricing density'' one can find a concrete formulation of the optimisation problem for which the result does not hold.
Equivalently, one can preserve the result after discarding any of the preceding assumptions only under a ``collapse to the mean'':
The ``pricing density'' must be constant, and the ``pricing rule'' in the budget constraint can be expressed by a standard expectation.

\begin{proposition}
For every nonconstant $D\in\cX^\ast$ with $\E[D]>0$ there exists a feasible quadruple $(\ph,\cC,D,p)$ such that:
\begin{enumerate}[(i)]
\item $\ph$ is weakly increasing but no optimal solution is antimonotone with $D$.
\item $\cC$ is increasing but no optimal solution is antimonotone with $D$.
\item $\ph$ is increasing and $\Max(\ph,\cC,D,p)\in\R$ but there exist optimal solutions that are not antimonotone with $D$.
\item $\cC$ is increasing and $\Max(\ph,\cC,D,p)\in\R$ but there exist optimal solutions that are not antimonotone with $D$.
\end{enumerate}
\end{proposition}
\begin{proof}
Let $Z\in\cX$ be nonconstant and comonotone with $D$. Note that $Z$ is not antimonotone with $D$ due to Lemma~\ref{lem:interior}. Up to an appropriate translation, we can always assume that $\E[Z]=0$. Set $p=\E[DZ]$ and observe that $p>\E[D]\E[Z]=0$ again by Lemma~\ref{lem:interior}. We claim that there always exist a law-invariant functional $\ph$ and a law-invariant set $\cC$ such that $(\ph,\cC,D,p)$ is a feasible quadruple with the required properties and with respect to which $Z$ is an optimal solution.

First, consider the law-invariant set $\cC=\{X\in\cX \,; \ \E[X]\leq 0\}$ and set for every $X\in\cX$
\[
\ph(X)=\E[X].
\]
Clearly, $\ph$ is both weakly increasing and increasing. Note that $(\ph,\cC,D,p)$ is a feasible quadruple and $Z$ is an optimal solution with $\ph(Z)\in\R$. This shows (iii). In addition, by Lemma~\ref{lem:interior}, any optimal solution $X\in\cX$ that is antimonotone with $D$ would need to satisfy
\[
0<p=\E[DX]\le\E[D]\E[X]=\E[D]\E[Z]=0,
\]
which is clearly impossible. This shows that (i) holds.

Next, consider the law-invariant set $\cC=\{Z'+m \,; \ Z'\sim Z, \ m\in\R\}$ and set for every $X\in\cX$
\[
\ph(X)=
\begin{cases}
-|\E[X]|& \mbox{if} \ X\in\cC,\\
\infty & \mbox{otherwise}.
\end{cases}
\]
Clearly, $\cC$ is increasing. Note that $(\ph,\cC,D,p)$ is a feasible quadruple and $Z$ is an optimal solution with $\ph(Z)\in\R$. This shows that (iv) holds. In addition, by Lemma~\ref{lem:interior}, any optimal solution $X\in\cX$ that is antimonotone with $D$ would have to satisfy
\[
0<p=\E[DX]\le\E[D]\E[X]=\E[D]\E[Z]=0,
\]
which is clearly impossible. This shows that (ii) holds.
\end{proof}

We strengthen the previous result in two ways. In a first step, we show that imposing no condition on the domain $\cC$ besides law invariance leads to counterexamples independently of the choice of both the ``pricing density'' $D$ and the objective function $\ph$.

\begin{proposition}\
\begin{enumerate}[(i)]
    \item For every law-invariant $\ph\colon\cX\to[-\infty,\infty]$ and for every nonconstant $D\in\cX^\ast$ with $\E[D]>0$ there exists a feasible quadruple $(\ph,\cC,D,p)$ such that no optimal solution is antimonotone with $D$.
    \item For every law-invariant $\ph\colon\cX\to[-\infty,\infty]$ such that $\ph(X)\in\R$ for some nonconstant $X\in\cX$ and for every nonconstant $D\in\cX^\ast$ with $\E[D]>0$ there exists a feasible quadruple $(\ph,\cC,D,p)$ such that $\Max(\ph,\cC,D,p)\in\R$, but there exist optimal solutions that are not antimonotone with $D$.
\end{enumerate}
\end{proposition}
\begin{proof}
To show (i), take any nonconstant $Z\in\cX$ that is comonotone with $D$ and set $p=\E[DZ]$. In addition, set $\cC=\{Z'\in\cX \,; \ Z'\sim Z\}$. It is clear that $\cC$ is law invariant and that $(\ph,\cC,D,p)$ is a feasible quadruple with respect to which $Z$ is optimal. If $X\in\cX$ is another optimal solution, then we must have $X\sim Z$ as well as $\E[DX]=\E[DZ]$. As $Z$ is nonconstant, it follows from Lemma~\ref{lem:interior} that $X$ cannot be antimonotone with $D$. To show (ii), it suffices to repeat the same argument under the additional condition that $\ph(Z)\in\R$, which is possible by assumption.
\end{proof}

We reinforce the same message by showing that the monotonicity assumption on $\cC$ remains critical even if we impose more structure on the set $\cC$ itself. We illustrate this by focusing on two common choices in the literature, starting from an ``interval-like'' set.

\begin{proposition}
Let $\cC\subset\cX$ be law invariant and such that
\[
\cC = \{X\in\cX \,; \ a\leq X\leq b\}
\]
for suitable constants $a<b$. For every nonconstant $D\in\cX^\ast$ with $\E[D]>0$ there exists a feasible quadruple $(\ph,\cC,D,p)$ such that:
\begin{enumerate}[(i)]
\item $\ph$ is weakly increasing but no optimal solution is antimonotone with $D$.
\item $\ph$ is increasing and $\Max(\ph,\cC,D,p)\in\R$ but there exist optimal solutions that are not antimonotone with $D$.
\end{enumerate}
\end{proposition}
\begin{proof}
By assumption on $D$, we find $k\in\R$ such that $\probp(D\leq k)\in(0,1)$ and $\E[D\ind_{\{D\le k\}}]\neq 0$. Define for every $X\in\cX$
\[
\ph(X) = \frac{1}{\P(D>k)}\int_{\probp(D\leq k)}^1q_X(s)ds.
\]
Note that $\ph$ is both weakly increasing and increasing. Indeed, for all $X\in\cX$ and $m>0$ we have $\ph(X)\in\R$ and $\ph(X+m) = \ph(X)+m> \ph(X)$. Now, set
\[
Z=a\ind_{\{D\leq k\}}+b\ind_{\{D>k\}}\in\cC
\]
as well as $p=\E[DZ]$. Note that $Z$ is not constant and satisfies $\ph(X)\leq b=\ph(Z)$ for every $X\in\cC$. As a result, $(\ph,\cC,D,p)$ is a feasible quadruple and $Z$ is an optimal solution. Since, by construction, $Z$ is not antimonotone with $D$, we infer that (ii) holds. In addition, take any optimal solution $X\in\cX$ that is antimonotone with $D$. From $X\le b$ and
\[
\ph(X)=\ph(Z)=b,
\]
we infer that $q_X(s)=b$ for almost every $s\in[\P(D\le k),1)$. Consequently, $q_X(s)=b$ holds for almost every $s\in (0,\P(D\le k)]$ as well by antimonotonicity. As a result, we must have $X=b$, from which we deduce \[
a\E[D\ind_{\{D\leq k\}}]+b\E[D\ind_{\{D>k\}}]=\E[DZ]=\E[DX]=b\E[D].
\]
Hence, $\E[D\ind_{\{D\le k\}}]=0$, a contradiction to the choice of $k$. To avoid this contradiction, $D$ has to be constant. This shows that (i) holds.
\end{proof}

We conclude by focusing on the situation where $\cC$ admits a maximum with respect to a suitable preference relation. Recall that a binary relation $\succeq$ on $\cX$ is a {\em preference} if it is reflexive and transitive. A preference is {\em compatible with the expectation} if for all $X,Y\in\cX$
\[
X\succeq Y \ \implies \ \E[X]\geq\E[Y].
\]
This weak compatibility property is satisfied by many preference relations encountered in the literature, including the convex order and second-order stochastic dominance.

\begin{proposition}
Let $\cC\subset\cX$ be law invariant and such that, for a suitable $B\in\CC$ and a preference $\succeq$ compatible with the expectation,
\[
\CC\subset\{Y\in\CX\,;\ Y\preceq B\}.
\]
For every nonconstant $D\in\cX^\ast$ with $\E[D]>0$ there exists a feasible quadruple $(\ph,\cC,D,p)$ such that:
\begin{enumerate}[(i)]
\item $\ph$ is weakly increasing but no optimal solution is antimonotone with $D$.
\item $\ph$ is increasing and $\Max(\ph,\cC,D,p)\in\R$ but there exist optimal solutions that are not antimonotone with $D$.
\end{enumerate}
\end{proposition}
\begin{proof}
Let $Z\sim B$ be comonotone with $D$. Set $p=\E[DZ]$ and define for every $X\in\cX$
\[
\ph(X) = \E[X].
\]
Clearly, $\ph$ is both weakly increasing and increasing.
Note that $(\ph,\cC,D,p)$ is a feasible quadruple with respect to which $Z$ is an optimal solution with $\ph(Z)\in\R$.
As $Z$ is nonconstant and comonotone with $D$, it follows from Lemma~\ref{lem:interior} that $Z$ is not antimonotone with $D$, showing (ii). In addition, take any optimal solution $X\in\cX$ that is antimonotone with $D$. If $X$ were nonconstant, then we would derive from Lemma~\ref{lem:interior} that
\[p=\E[DX]<\E[D]\E[X]=\E[D]\E[Z]<\E[DZ]=p,\]
which is absurd. Hence, $X$ must be constant and equal to $\frac{p}{\E[D]}$ or equivalently $\frac{\E[DZ]}{\E[D]}$. By optimality and compatibility with the expectation, $X\in\cC$ yields
\[
\frac{\E[DZ]}{\E[D]} = \E[X] \leq \E[B] = \E[Z].
\]
This implies $\E[DZ]\leq\E[D]\E[Z]$, which is, however, in contrast to the comonotonicity between $Z$ and $D$ by Lemma~\ref{lem:interior}. This shows that (i) holds.
\end{proof}


\begin{appendix}
\section{Proof of Lemma~\ref{lem:interior}}\label{appendix}

\begin{proof}[Proof of Lemma~\ref{lem:interior}]
First, let $X$ and $Y$ be positive. For every $X'\sim X$, Fubini's theorem yields
\begin{align*}
\E[X'Y]
&=
\E\left[\int_0^\infty\int_0^\infty\ind_{[0,X')}(x)\ind_{[0,Y)}(y)dxdy\right] \\
&=
\int_0^\infty\int_0^\infty\E[\ind_{\{X'>x\}}\ind_{\{Y>y\}}]dxdy \\
&=
\int_0^\infty\int_0^\infty\probp(X'>x,Y>y)dxdy \\
&\leq
\int_0^\infty\int_0^\infty\min\{\probp(X'>x),\probp(Y>y)\}dxdy \\
&=
\int_0^\infty\int_0^\infty\int_0^1\ind_{[F_X(x),1]}(s)\ind_{[F_Y(y),1]}(s)dsdxdy \\
&=
\int_0^1\int_0^{q_Y(s)}\int_0^{q_X(s)}dxdyds
=
\int_0^1q_X(s)q_Y(s)ds.
\end{align*}
We have equality if and only if $\probp(X'>x,Y>y)=\min\{\probp(X'>x),\probp(Y>y)\}$, or equivalently $\probp(X'\leq x,Y\leq y)=\min\{\probp(X'\leq x),\probp(Y\leq y)\}$, for almost all $x,y\in\R$ with respect to the Lebesgue measure on $\R\times\R$. By right continuity of distribution functions, this holds if and only if $X'$ and $Y$ are comonotone. Note that, by Lemma~\ref{lem:existence}, we do find $X'\sim X$ such that $X'$ and $Y$ are comonotone. This proves the integrability of $q_Xq_Y$, the right-hand side equality in \eqref{eq:interval}, and the corresponding attainability assertion. In a similar way, we obtain
\begin{align*}
\E[X'Y]
&=
\int_0^\infty\int_0^\infty\probp(X'>x,Y>y)dxdy \\
&\geq
\int_0^\infty\int_0^\infty\max\{\probp(X'>x)-\probp(Y\leq y),0\}dxdy \\
&=
\int_0^\infty\int_0^\infty\int_0^1\ind_{[0,1-F_X(x)]}(s)\ind_{[F_Y(y),1]}(s)dsdxdy \\
&=
\int_0^1\int_0^{q_Y(s)}\int_0^{q_X(1-s)}dxdyds
=
\int_0^1q_X(1-s)q_Y(s)ds.
\end{align*}
We have equality if and only if $\probp(X'>x,Y>y)=\max\{\probp(X'>x)-\probp(Y\le y),0\}$, or equivalently $\probp(X'\leq x,Y\leq y)=\max\{\probp(X'\leq x)+\probp(Y\leq y)-1,0\}$, for almost all $x,y\in\R$ with respect to the Lebesgue measure on $\R\times\R$. By right continuity of distribution functions, this holds if and only if $X'$ and $Y$ are antimonotone. Note that, by Lemma~\ref{lem:existence}, we do find $X'\sim X$ such that $X'$ and $Y$ are antimonotone. This proves the integrability of $q_X(1-\cdot)q_Y$, the left-hand side equality in \eqref{eq:interval}, and the corresponding attainability assertion. The statement for general $X$ and $Y$ follows by applying \eqref{eq:interval} and the attainability result to the positive and negative parts of $X$ and $Y$ exploiting the fact that $q_{\max\{X,0\}}=\max\{q_X,0\}$ and $q_{\max\{-X,0\}}=\max\{-q_X(1-\cdot),0\}$ almost surely with respect to the Lebesgue measure on $(0,1)$, and similarly for $Y$. For the attainability assertion,
 one observes that $X$ and $Y$ are comonotone if and only if $\max\{X,0\}$ and $\max\{Y,0\}$ as well as $\max\{-X,0\}$ and $\max\{-Y,0\}$ are comonotone and $\max\{X,0\}$ and $\max\{-Y,0\}$ as well as $\max\{-X,0\}$ and $\max\{Y,0\}$ are antimonotone, and similarly for antimonotonicity.

Now, take general nonconstant $X$ and $Y$. Observe that
\begin{align*}
&2\left(\int_0^1q_X(s)q_Y(s)ds-\E[X]\E[Y]\right) \\
&= \int_0^1\int_0^1q_X(s)q_Y(s)dtds+\int_0^1\int_0^1q_X(t)q_Y(t)dtds-2\int_0^1\int_0^1q_X(s)q_Y(t)dtds\\
&=\int_0^1\int_0^1\big(q_X(s)-q_X(t)\big)\big(q_Y(s)-q_Y(t)\big)dtds.
\end{align*}
The integrand in the last expression is nonnegative. Moreover, we can invoke nonconstancy of $X$ and $Y$ to find some $\alpha\in(0,\tfrac 1 2)$ such that $q_X(t)-q_X(s)>0$ and $q_Y(t)-q_Y(s)>0$ for all $s<\alpha$ and $t>1-\alpha$. This shows the right-hand side inequality in~\eqref{eq: strict hardy littlewood}. Repeating the argument by replacing $X$ with $-X$ delivers the left-hand side inequality in~\eqref{eq: strict hardy littlewood} and concludes the proof.
\end{proof}

\begin{remark}
The strict inequality in \eqref{eq: strict hardy littlewood} is seldom found in the literature and is related to a rearrangement inequality by Chebyshev; see, e.g., \cite{Chebyshev}. An alternative proof can be obtained from \cite[Lemma 8]{Xu}. Indeed, by nonatomicity of $(\Omega,\CF,\P)$ we find two independent random variables $U_1$ and $U_2$ with uniform distribution over $(0,1)$. Hence, $X':=q_X(U_1)\sim X$ and $Y':=q_Y(U_2)\sim Y$ are independent as well. Let $\alpha\in (0,\tfrac 1 2)$ be such that $q_X(s)<q_X(t)$ and $q_Y(s)<q_Y(t)$ for all $s\le \alpha$ and $t\ge 1-\alpha$, which is possible as $X$ and $Y$ are not constant. Set
\[
R=(\{U_1\ge1-\alpha\}\cap\{U_2\le \alpha\})\times(\{U_1\le\alpha\}\cap\{U_2\ge1- \alpha\})
\]
and note that $(\P\otimes\P)(R)=\alpha^4>0$ and that
\[
\Omega\times\Omega\ni(\om,\om')\mapsto\left(X'(\om)-X'(\om')\right)\cdot\left(Y'(\om)-Y'(\om')\right)
\]
is negative $\P\otimes\P$-almost surely on $R$. As the random variables $X'$ and $Y'$ can therefore not be comonotone, we obtain
\[
\E[X]\E[Y]=\E[X'Y']<\E[q_X(U_1)q_Y(U_1)]=\int_0^1q_X(s)q_Y(s)ds.
\]
The other inequality follows by exchanging $X$ with $-X$.
\end{remark}

\end{appendix}


\end{document}